\documentclass{PoS}

\newcommand{\lsim}{
\mathrel{\hbox{\rlap{\hbox{\lower4pt\hbox{$\sim$}}}\hbox{$<$}}}}
\newcommand{\beq}{\begin{equation}}
\newcommand{\eeq}{\end{equation}}
\newcommand{\bea}{\begin{eqnarray}}
\newcommand{\eea}{\end{eqnarray}}

\title{The Standard model prediction for $K_{e2}/K_{\mu2}$ and $\pi_{e2}/\pi_{\mu 2}$}

\ShortTitle{The Standard model prediction for $K_{e2}/K_{\mu2}$ and $\pi_{e2}/\pi_{\mu 2}$}

\author{\speaker{Ignasi Rosell$^{ab}$}\,  \thanks{I wish to thank the organizers of CD09 for the useful and pleasant congress. 
 I want to thank V.~Cirigliano for their comments and suggestions. 
This work is supported in part by the Universidad CEU Cardenal Herrera (Grant No. PRCEU-UCH20/08), by the Spanish Government (Grant No. FPA2007-60323 and Consolider-Ingenio 2010 No. CSD2007-00042, CPAN) and by the EU No. MRTN-CT-2006-035482 (FLAVIAnet).}
\\
 \llap{$^a$}Departamento de Ciencias F\'\i sicas, Matem\'aticas y de la Computaci\'on, Universidad CEU Cardenal Herrera, c/ Sant Bartomeu 55, E-46115 Alfara del Patriarca, Val\`encia, Spain \\
 \llap{$^b$}  IFIC, Universitat de Val\`encia - CSIC, Apartat de Correus 22085, E-46071 Val\`encia, Spain \\

        E-mail: \email{rosell@uch.ceu.es}}

\abstract{We have calculated the ratios $R_{e/\mu}^{(P)}  \equiv \Gamma( P \to e \bar{\nu}_e [\gamma] )/ \Gamma( P \to \mu \bar{\nu}_\mu [\gamma] )$    ($P=\pi,K$) in Chiral Perturbation Theory up to $\mathcal{O}(e^2p^4)$,  finding $R_{e/\mu}^{(\pi)} = ( 1.2352 \pm 0.0001 ) \times 10^{-4}$ and  $R_{e/\mu}^{(K)} = ( 2.477 \pm 0.001 ) \times 10^{-5}$. This observable is helicity suppressed in the Standard Model, so that it is a sensitive probe of all Standard Model extensions that induce pseudoscalar currents and nonuniversal corrections to the lepton couplings. Ongoing experimental searches plan to reach uncertainties that are comparable to these results. At the moment $R_{e/\mu}^{(K)}$ is in agreement with the final result by the KLOE Collaboration at DAFNE and it is at $1.4\,\sigma$ of the preliminary result by the NA62 Experiment at CERN. New measurements of $R_{e/\mu}^{(\pi)}$ are under way by the PEN Collaboration at PSI and by the PIENU Collaboration at TRIUMF.  }

\FullConference{6th International Workshop on Chiral Dynamics\\
                 July 6-10 2009\\
                 Bern, Switzerland}

\begin{document}

\section{Motivation}

The ratio $R_{e/\mu}^{(P)} \equiv  \Gamma( P \to e \bar{\nu}_e [\gamma] )/  \Gamma( P \to \mu \bar{\nu}_\mu [\gamma] )$    ($P=\pi,K$) of leptonic decay rates of light pseudoscalar mesons is  helicity-suppressed in the Standard Model, due to the $V-A$ charged current coupling. It is therefore a  sensitive probe of  all  Standard Model extensions that induce pseudoscalar currents and non-universal corrections to the lepton couplings~\cite{Bryman:1993gm}. Attention to these process has been payed in the context of the Minimal Supersymmetric Standard Model, with~\cite{Masiero:2005wr}  and without~\cite{RamseyMusolf:2007yb} lepton-flavor-violating effects. 
In general, effects from weak-scale new physics  are expected in the range  $(\Delta R_{e/\mu})/R_{e/\mu} \sim  10^{-4} - 10^{-2}$  and there is a realistic chance to detect or constrain them because of  the following circumstances:
\begin{enumerate}
\item[{\it i})] Ongoing experimental searches plan to reach a fractional uncertainty of $(\Delta R^{(\pi)}_{e/\mu})/R^{(\pi)}_{e/\mu} \sim  5 \times  10^{-4}$~\cite{PEN,PIENU} and  $(\Delta R^{(K)}_{e/\mu})/R^{(K)}_{e/\mu} \sim 3 \times 10^{-3}$~\cite{KLOE,NA62}, which represent respectively a factor of  around $5$  and $10$ improvement over former errors~\cite{Rp1,Rp2,Rp3,RK1,RK2,RK3}. 
\item[{\it ii})] At the same time, the Standard Model theoretical uncertainty  can be pushed below this level,  since to a first approximation the strong interaction dynamics cancels out in the ratio $R_{e/\mu}^{(P)}$ and hadronic structure dependence  appears only through electroweak corrections.   Indeed, the most recent  theoretical predictions read  $R^{(\pi)}_{e/\mu} = (1.2352   \pm  0.0005) \times 10^{-4}$~\cite{MS93},  $R^{(\pi)}_{e/\mu} = (1.2354   \pm  0.0002) \times 10^{-4}$~\cite{Fink96}, and $R^{(K)}_{e/\mu} = (2.472   \pm  0.001) \times 10^{-5}$~\cite{Fink96}. In Ref.~\cite{MS93}  a general parameterization of the hadronic effects is given, with an estimate of the leading model-independent contributions based on current algebra~\cite{terentev}. The dominant  hadronic  uncertainty is  roughly estimated  via  dimensional analysis. In Ref.~\cite{Fink96}, on the other hand,  the hadronic component is calculated by modeling the low- and intermediate-momentum region of the loops involving virtual photons.  
\end{enumerate}

\section{The Standard Model prediction}

In Refs.~\cite{PRL,JHEP} we have  analyzed $R_{e/\mu}^{(P)}$   within  Chiral Perturbation Theory~\cite{chpt}, the low-energy effective field theory of QCD.  The key feature of this  framework is that it provides  a controlled expansion of the amplitudes  in terms of the masses of pseudoscalar mesons and charged leptons  ($p \sim m_{\pi,K, \ell}/\Lambda_\chi$, with $\Lambda_\chi \sim 4 \pi F_\pi \sim 1.2 \,  {\rm GeV}$),  and the electromagnetic  coupling ($e$). Electromagnetic corrections to (semi)-leptonic decays of   $K$ and $\pi$ have been worked out to  $\mathcal{O}(e^2 p^2)$~\cite{Knecht:1999ag,Semileptonic}, but had never been pushed  to  $\mathcal{O}(e^2 p^4)$, as  required for $R_{e/\mu}^{(P)}$  in order to match the experimental accuracy.

Within the chiral power counting,  $R_{e/\mu}$ is  written as:
\bea
R_{e/\mu}^{(P)} &= & R_{e/\mu}^{(0),(P)} \left( 1+\Delta_{LL} \right)  \, \Bigg[   1 +  \Delta_{e^2 p^2}^{(P)} +   \Delta_{e^2 p^4}^{(P)}  + \Delta_{e^2 p^6}^{(P)}   +  ... \Bigg] \,, \label{master}
\eea 
being $R_{e/\mu}^{(0),(P)}$ the well known tree-level expression:
\bea
R_{e/\mu}^{(0),(P)} & =& \frac{m_e^2}{m_\mu^2}  \left(  \frac{m_P^2 - m_e^2}{m_P^2 - m_\mu^2}  \right)^2  ~.  \label{eq:R0}
\eea
At the level of uncertainty considered here, one needs to include higher order long distance corrections~\cite{MS93} and their effect amounts to the factor $1+\Delta_{LL}$ in (\ref{master}),
\beq
1 + \Delta_{LL} = 
\displaystyle\frac{\left(1 - \frac{2}{3} \frac{\alpha}{\pi} \log \frac{m_\mu}{m_e} \right)^{9/2}}{1 -
 \frac{3 \alpha}{\pi} \log \frac{m_\mu}{m_e}} = 
1.00055~.
\eeq
The leading electromagnetic correction  $\Delta_{e^2 p^2}^{(P)} $ corresponds to  the point-like approximation for pion and kaon, and its expression is also well known~\cite{MS93,Knecht:1999ag,Kinoshita:1959ha}:
\bea
\Delta^{(P)}_{e^2 p^2} &= & \frac{\alpha}{\pi}  \Big[ F( \frac{m_e^2}{m_P^2}) - F(\frac{m_\mu^2}{m_P^2})  \Big] \,,  \\
F(z)   &=&  \frac{3}{2} \log z  + \frac{13 - 19 z}{8 ( 1 - z)} - 
\frac{8 - 5 z}{4 (1 - z)^2}  \, z   \log z - \left(  2 + \frac{1 + z}{1 - z} \log z  \right)  \log ( 1 - z)  \nonumber \\
&-& 2 \frac{1 + z}{1 - z}  \, Li_2  ( 1 - z)     ~.
\eea
The structure dependent effects are all contained in $\Delta_{e^2 p^4}^{(P)}$ and higher order terms,  which are the main subject of Refs.~\cite{PRL,JHEP}. Neglecting terms of order $(m_e/m_\rho)^2$,  the most general parameterization of the next-to-leading  chiral contribution can be written in the form 
\beq
 \Delta_{e^2 p^4}^{(P)} = \frac{\alpha}{\pi} \frac{m_\mu^2}{m_\rho^2}  \left(c_2^{(P)}  \, \log \frac{m_\rho^2}{m_\mu^2}  
+  c_3^{(P)}  + c_4^{(P)} (m_\mu/m_P) \right) +  \frac{\alpha}{\pi}  \frac{m_P^2}{m_\rho^2} \,  \tilde{c}_{2}^{(P)}  \, \log \frac{m_\mu^2}{m_e^2} ~ , \label{eq:dele2p4}
\eeq
\begin{table}[t!]
\begin{center}
\begin{tabular}{|c|c|c|}
\hline
  & $(P=\pi)$  & $(P=K)$    \\[5pt]
\hline
 $\tilde{c}_2^{(P)}$  &   0  &  $ (7.84 \pm 0.07_\gamma) \times 10^{-2}  $ \\
 $c_2^{(P)}$  & $5.2 \pm 0.4_{L_9} \pm 0.01_\gamma$  &  $4.3 \pm 0.4_{L_9} \pm 0.01_\gamma $ \\
 $c_3^{(P)}$   
&   
$ -10.5 \pm 2.3_{\rm m } \pm 0.53_{L_9} 
$
& 
$ -4.73 \pm 2.3_{ \rm m} \pm 0.28_{L_9}$ 
\\
 $c_4^{(P)} (m_\mu) $  &
$1.69 \pm 0.07_{L_9} $
&  
$ 0.22 \pm 0.01_{L_9} $ 
\\
\hline
\end{tabular}
\end{center}
\caption{Numerical values for $c_{2,3,4}^{(P)}$ and $\tilde{c}_2^{(P)}$, for $P=\pi,K$. The uncertainties correspond to the input values  $L_9^r (\mu=m_\rho) = (6.9 \pm 0.7) \times 10^{-3} $, $\gamma= 0.465 \pm 0.005$~\cite{pocanic}, and to the matching procedure (${\mathrm m}$),  affecting only $c_3^{(P)}$.}
\label{tab:tab1}
\end{table}
which highlights the dependence on lepton masses. The dimensionless constants  $c_{2,3}^{(P)}$  do not  depend on the lepton mass 
but depend logarithmically on hadronic masses, while  $c_4^{(P)} (m_\mu/m_P) \to 0$ as $m_\mu \to 0$. (Note that  our $c_{2,3}^{(\pi)}$  do not coincide with $C_{2,3}$ of Ref.~\cite{MS93}, because their $C_{3}$ is not constrained to be $m_\ell$-independent and contains in general logarithms of $m_\ell$.)  
  
Let us note that the  results for $c_{2,3,4}^{(P)}$ and $\tilde{c}_{2}^{(P)}$   depend on the definition of the inclusive rate  $\Gamma (P \to \ell \bar{\nu}_\ell [\gamma])$. The radiative  amplitude is the sum of the  inner bremsstrahlung ($T_{IB}$) component of $\mathcal{O}(e p)$ and a structure dependent ($T_{SD}$) component of $\mathcal{O}(e p^3)$~\cite{Bijnens:1992en}. The experimental  definition of $R_{e/\mu}^{(\pi)}$ is fully inclusive on the radiative mode, so that $\Delta_{e^2 p^4}^{(\pi)}$ receives a contribution  from the interference
of $T_{IB}$ and $T_{SD}$.  Moreover, in this case one also has to include the effect of $\Delta_{e^2 p^6}^{(\pi)} \propto |T_{SD}|^2$, 
that is formally of $\mathcal{O}(e^2 p^6)$, but is not helicity suppressed and behaves as $\Delta_{e^2 p^6} \sim   \alpha/\pi  \,  (m_P/M_V)^4  \, (m_P/m_e)^2$. On the other hand, the usual experimental definition of  $R_{e/\mu}^{(K)}$ is not fully inclusive on the radiative mode.  It corresponds to including the effect of $T_{IB}$ in $\Delta_{e^2 p^2}^{(K)}$ (dominated by soft photons) and excluding altogether the effect of $T_{SD}$: consequently $c_n^{(\pi)} \neq c_n^{(K)}$. 

The expressions of the constants $c_{2,3,4}^{(P)}$ and $\tilde{c}_2^{(P)}$ are shown in Refs.~\cite{PRL,JHEP} and their numerical values are reported in table~\ref{tab:tab1}. Note that to this order in Chiral Perturbation Theory, $R_{e/\mu}^{(P)}$ features both model independent double chiral logarithms (previously neglected) and an a priori unknown low-energy  constant. By including  the finite loop effects and estimating the low-energy constant via a matching calculation in large-$N_C$ QCD,  we thus provide the first complete  result  of $R_{e/\mu}^{(P)}$ to $\mathcal{O}(e^2 p^4)$ in  the effective power counting.  Most importantly, the matching calculation allows us to  further reduce the theoretical uncertainty and put it on more solid ground.

In table~\ref{tab:tab2} we summarize the various electroweak corrections to  $R_{e/\mu}^{(\pi,K)}$.  Applying these  we arrive to our final results:
\bea
R_{e/\mu}^{(\pi)} &=& (1.2352 \pm 0.0001 ) \times 10^{-4} \,,
\label{eq:Rpf} \\ 
R_{e/\mu}^{(K)} &=& ( 2.477 \pm 0.001 ) \times 10^{-5} ~ . 
\label{eq:RKf}
\eea
The uncertainty we quote for $R_{e/\mu}^{(\pi)}$  is entirely induced by our matching procedure. However, in the case of $R_{e/\mu}^{(K)}$ we have inflated the nominal uncertainty arising from matching  by a factor of four,  to account for higher order chiral corrections, that  are expected  to scale as  $\Delta_{e^2 p^4}^{(K)} \times m_K^2/(4 \pi F)^2$.   
\begin{table}[t!]
\begin{center}
\begin{tabular}{|c|c|c|}
\hline
 & $(P=\pi)$  & $(P=K)$    \\[5pt]
\hline
 $\Delta_{e^2 p^2}^{(P)} \ \, (\%)  $   &   $-3.929 $ &  $ -3.786 $ \\
$\Delta_{e^2 p^4}^{(P)}  \ \, (\%) $ 
 & $0.053 \pm 0.011$  &  $0.135 \pm 0.011$ \\
$\Delta_{e^2 p^6}^{(P)} \ \,  (\%)  $ 
&   
$ 0.073$
& 
\\
$\Delta_{LL} \ \ (\%) $   &
$ 0.055 $ 
&  
$ 0.055 $ 
\\
\hline
\end{tabular}
\end{center}
\caption{Numerical summary of various electroweak corrections to $R_{e/\mu}^{(P)}$. 
The uncertainty in $\Delta_{e^2 p^4}^{(P)}$ corresponds to the matching procedure. }
\label{tab:tab2}
\end{table}

\section{Discussion}

\subsection{Comparison to previous theoretical predictions}

Our results have to be compared with the previous theoretical predicitions of Refs.~\cite{MS93} and \cite{Fink96}, which we report in table~\ref{tab:tab3}. 
\begin{enumerate}
\item[{\it i})] $R_{e/\mu}^{(\pi)}$ is  in good agreement with both previous results.
\item[{\it ii})] There is a discrepancy  in $R_{e/\mu}^{(K)}$ that goes well outside the estimated theoretical uncertainties. We have traced back  this difference  to  two problematic aspects of  Ref.~\cite{Fink96}. The leading log correction $\Delta_{LL}$ is included  with the wrong sign: this accounts for half  of the discrepancy. The remaining effect is due to the difference in the next-to-leading order virtual correction, for which  Finkemeier finds $\Delta_{e^2 p^4}^{(K)}  =  0.058 \%$. We have serious doubts on the reliability of this number  because the hadronic form factors 
modeled in Ref.~\cite{Fink96}  do not satisfy the correct QCD short-distance behavior. At high momentum they  fall off faster than the QCD requirement, thus leading to a smaller value of  $\Delta_{e^2 p^4}^{(K)}$ compared to our work. 
\end{enumerate}
\begin{table}[t!]
\begin{center}
\begin{tabular}{|c|c|c|}
\hline
 & $10^4 \cdot R_{e/\mu}^{(\pi)}$  & $ 10^5 \cdot  R_{e/\mu}^{(K)}$    \\[5pt]
\hline
 This work   &    $1.2352 \pm 0.0001$ & 
$ 2.477 \pm 0.001 $ \\
Ref.~\cite{MS93}   & 
$1.2352 \pm 0.0005$ 
 & 
 \\
Ref.~\cite{Fink96}
&   
$1.2354 \pm 0.0002$ 
&  
$ 2.472 \pm 0.001 $ 
\\
\hline
\end{tabular}
\end{center}
\caption{Comparison of our result with previous theoretical predictions of  $R_{e/\mu}^{(P)}$.}
\label{tab:tab3}
\end{table}

\subsection{Comparison to experiments}

\subsubsection{$R_{e/\mu}^{(\pi)}$}

The three most recent measurements of $R_{e/\mu}^{(\pi)}$ are mutually consistent:
\bea
R_{e/\mu}^{(\pi)}|_{\mathrm{Bryman}}&=&\left( 1.218 \pm 0.014 \right) \times 10^{-4} \quad \mathrm{Ref.~\cite{Rp1}}\,, \nonumber \\
R_{e/\mu}^{(\pi)}|_{\mathrm{Britton}}&=& \left( 1.2265 \pm 0.0034_{\mathrm{stat}} \pm 0.0044_{\mathrm{syst}} \right) \times 10^{-4} \,=\,  \left( 1.227 \pm 0.006 \right) \times 10^{-4} \quad \mathrm{Ref.~\cite{Rp2}}\,, \nonumber \\
R_{e/\mu}^{(\pi)}|_{\mathrm{Czapek}}&=& \left( 1.2346 \pm 0.0035_{\mathrm{stat}} \pm 0.0036_{\mathrm{syst}} \right) \times 10^{-4} \,=\,\left( 1.235 \pm 0.005\right) \times 10^{-4}\quad \mathrm{Ref.~\cite{Rp3}}\,. \nonumber \\ 
\eea
These measurements are in agreement with the theoretical prediction and they give the PDG average $\left( 1.230 \pm 0.004 \right) \times 10^{-4}$~\cite{PDG}. Note that it is less accurate than the prediction by a factor of around 40. As it has been indicated previously, the experiments ruled by the PEN Collaboration at the Paul Scherrer Institute~\cite{PEN}  and by the PIENU Collaboration at  TRIUMF~\cite{PIENU} are under way and are expected to improve significantly the uncertainty.

\subsubsection{$R_{e/\mu}^{(K)}$}

\begin{figure}[!t]
\begin{center}
\includegraphics[scale=0.42]{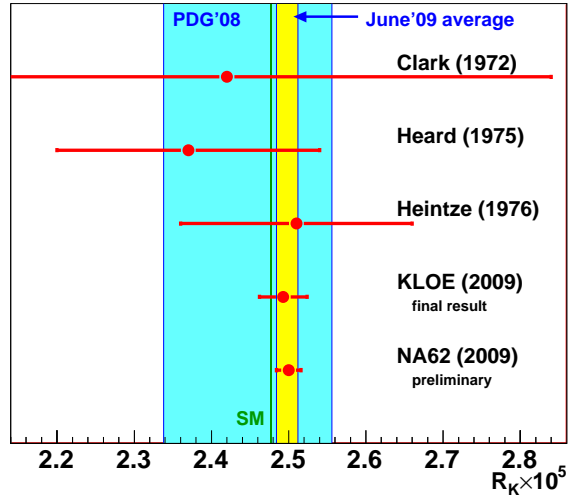}
\caption{ 
Summary of $R_{e/\mu}^{(K)}$ measurements (picture from \cite{NA62}). \label{fig:fig1}
}
\end{center}
\end{figure}

The old measurements of $R_{e/\mu}^{(K)}$ in the seventies are also mutually consistent and in agreement with the Standard Model prediction,
\bea
R_{e/\mu}^{(K)}|_{\mathrm{Clark}}&=&  \left( 2.42 \pm 0.42 \right) \times 10^{-5} \qquad \mathrm{Ref.~\cite{RK1}}\,, \nonumber \\
R_{e/\mu}^{(K)}|_{\mathrm{Heard}}&=& \left( 2.37 \pm 0.17\right) \times 10^{-5}\qquad \mathrm{Ref.~\cite{RK2}}\,, \nonumber \\ 
R_{e/\mu}^{(K)}|_{\mathrm{Heintze}}&=& \left( 2.51 \pm 0.15\right) \times 10^{-5}\qquad \mathrm{Ref.~\cite{RK3}}\,. \nonumber \\ 
\eea
Using these data the world average reads $\left( 2.45 \pm 0.11 \right) \times 10^{-5}$~\cite{PDG}. Again it is much less accurate than our theoretical prediction, by a factor of around 100. Recent experiments have been improving significantly the uncertainty:
\begin{enumerate}
\item [{\it i})] The KLOE Collaboration has recently published the measurement performed at DAFNE~\cite{KLOE},
\bea
R_{e/\mu}^{(K)}|_{\mathrm{KLOE}}&=& \left( 2.493 \pm 0.025_{\mathrm{stat}} \pm 0.019_{\mathrm{syst}} \right) \times 10^{-5}\,=\, \left( 2.49 \pm 0.03 \right) \times 10^{-5}  \,,
\eea
in agreement with our Standard Model prediction of \ref{eq:RKf}. The achieved precision is $1.2\%$, improving the precision of the former world average by a factor of 4.
\item [{\it ii})] On the other hand the NA62 experiment at CERN has also recently announced its preliminary result~\cite{NA62},
\bea
R_{e/\mu}^{(K)}|_{\mathrm{NA62}}&=& \left( 2.500 \pm 0.012_{\mathrm{stat}} \pm 0.011_{\mathrm{syst}} \right) \times 10^{-5}\,=\, \left( 2.500 \pm 0.016 \right) \times 10^{-5}  \,. 
\eea
The uncertainty is now $0.64\%$. Note that the whole 2007-08 data sample is supposed to allow pushing the uncertainty down to $0.4\%$. This result is compatible with the KLOE one and it is also in agreement with the theoretical prediction of  \ref{eq:RKf} (at $1.4~\sigma$).
\end{enumerate}
With these new results, and until the final result of NA62 arrives, using a simple weighted mean the new world average reads 
\bea
R_{e/\mu}^{(K)}|_{\mathrm{WA}}&=&\left( 2.498 \pm 0.014 \right) \times 10^{-5} \,,
\eea
which is in agreement with the Standard Model result (at $1.5~\sigma$). A summary of $R_{e/\mu}^{(K)}$ measurements is presented in figure~\ref{fig:fig1}.

\end{document}